# *In situ* Measurement of Biaxial Modulus of Si Anode for Li-ion Batteries

V.A. Sethuraman,[*] M.J. Chon, M. Shimshak, N. Van Winkle, P.R. Guduru[**]

School of Engineering, Brown University
Providence, Rhode Island 02912, USA

We report *in situ* measurement of biaxial moduli of a Si thin-film electrode as a function of its lithium concentration. During lithiation, biaxial compressive stress is induced in the Si film and it undergoes plastic flow. At any state-of-charge (SOC), a relatively small delithiation-relithiation sequence unloads and reloads the film elastically. From the stress and strain changes during a delithiation-relithiation cycle, the biaxial modulus of the film is calculated. Stress change is obtained by measuring the change in substrate curvature using a Multi-beam Optical Sensor; the elastic strain change is obtained from the change in SOC. By repeating these measurements at several different values of SOC, the biaxial modulus was seen to decrease from *ca.* 70 GPa for $Li_{0.32}Si$ to *ca.* 35 GPa for $Li_{3.0}Si$. Such a significant reduction in elastic modulus has important implications for modeling stress evolution and mechanical degradation in Si-based anodes.

[*] - vj@cal.berkeley.edu; +1 (510) 764-4842 (V.A. Sethuraman)
[**] - Pradeep_Guduru@Brown.edu; +1 (401) 863- 3362 (P.R. Guduru)

## 1. INTRODUCTION

Silicon's high lithiation capacity of 3579 mAh g$^{-1}$ [1] makes it an attractive choice for use as negative electrodes in lithium-ion batteries. A large body of literature exists on the electrochemical and mechanical performance of anodes made of pure silicon and composites in which silicon is one of the constituents [2]. It is well known that Si undergoes large volume expansion (*ca.* 370%) upon complete lithiation, which results in generation of stresses as high as 1-2 GPa [3]. The large stresses lead to mechanical damage and consequently, capacity fading. In order to model the mechanics of electrode materials and predict their cycle life, it is essential to know their mechanical properties (elastic moduli, flow stress and its rate dependence, ductility, fracture toughness, *etc*) and how they vary with SOC. When Si is fully lithiated, the number of solute atoms is about 3.5 times that of the solvent atoms; hence, it is reasonable to expect significant changes in mechanical properties with SOC. However, little data is available in literature on the mechanical properties of lithiated silicon. Recently, using density functional theory (DFT) calculations, Shenoy *et al.* estimated the elastic moduli of Li-Si system as a function of lithium concentration [4], and noted that they decrease linearly with increasing lithium concentration. For example, they predict that the Young's modulus of amorphous silicon decreases from about 90 GPa for α-Si to less than 40 GPa for α-Li$_{15}$Si$_4$. If the reduction in elastic moduli is also any indication of increased ductility due to alloying, one can expect significant changes in other mechanical properties as well. In this paper, we present an experimental effort whose objective is to directly measure the biaxial modulus of a silicon thin-film electrode as a function of Li concentration.

## 2. EXPERIMENTAL

*2.1. Electrochemical cell*

Si wafers [double-side polished, 50.8 mm diameter, nominally 425-450 μm thick, (111) orientation, with 200 nm thermal oxide on all sides] were used as substrates for electrode fabrication. The oxide layer isolates the silicon wafer from participating in the electrochemical reactions. A 50 nm Ti layer was first sputtered (Lesker Lab18 Sputtering System) on one of the sides of the Si wafer followed by deposition of *ca.* 250 nm thick Cu layer and *ca.* 250 nm thick silicon layer (see Figure 1). Film thicknesses were measured using a surface profilometer (Dektak) and a white-light interferometer (Zygo). Ti and Cu layers were prepared by DC-sputtering of Ti and Cu targets (2" diameter, 0.25" thick discs, 99.995%, Kurt Lesker Company) at 150 W and a pressure of 0.266 Pa of argon. The Ti layer improves adhesion between the wafer and the Cu layer. Previous studies have shown that the Cu underlayer is critical to the cycling of Si thin films [5]. Si thin films were prepared by RF-magnetron sputtering of a Si target (2" diameter, 0.25" thick disc, 99.995% Si) at 180 W power and at a pressure of 0.266 Pa of Ar (99.995%). Previous studies have shown that the sputter-deposited Si films are amorphous in nature [5]. The sample wafer (with the Ti, Cu and Si layers) is then assembled into an electrochemical cell, which is shown schematically in Figure 1, in ultra-high pure Ar atmosphere. Li metal (50.8 mm diameter, 1.5 mm thick) was used as counter and reference electrode, which is separated from the Si electrode by a woven Celgard C480 separator (62 mm diameter, 21.5 μm thick). 1.2 M lithium hexafluoro-phosphate in 3:7 (wt. %) ethylene carbonate: diethyl carbonate with 10% fluoro-ethylene carbonate was used as the electrolyte.

*2.2. Electrochemical measurements*

Electrochemical measurements were conducted in ultra-high pure Ar atmosphere at 25°C (±1°C) using a Solartron 1470E MultiStat (Solartron, Oak Ridge, TN). The Si thin-film electrode was lithiated galvanostatically at a current-density of 5 µA/cm$^2$ (*ca.* C/24 rate) for one hour followed by an open-circuit relaxation for one hour. The electrode was then delithiated galvanostatically at the same rate for 8 minutes followed by an open-circuit relaxation for 5 minutes (see Figure 2a for a representative sequence). These lithiation-relaxation-delithiation-relaxation-relithiation sequence of steps was repeated at several different values of SOC until a lower cut-off potential of 0.01 V *vs.* Li/Li$^+$ was reached.

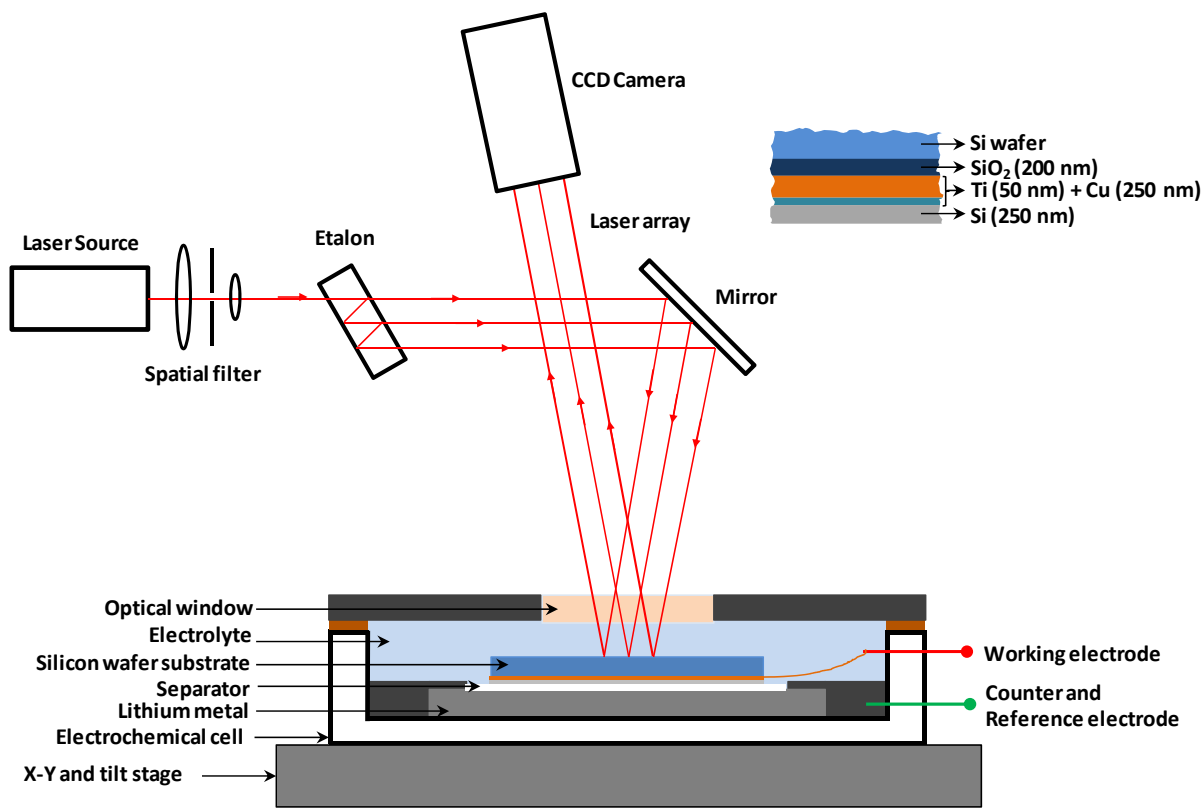

*Figure 1: Schematic illustration of the electrochemical cell and the MOS setup used to measure substrate curvature. In this two-electrode configuration, Si thin film is the working electrode, and Li metal is the counter and reference electrode. Above right: layered configuration of the working electrode on Si-wafer substrate is shown. Cell-schematic is not drawn to scale.*

*2.3. In situ stress measurements*

Stress in the Si film was measured by monitoring the substrate-curvature changes during electrochemical lithiation and delithiation. The relationship between the biaxial film stress, $\sigma$, and the substrate-curvature change $\Delta\kappa$, is given by the Stoney equation [6,7],

$$\sigma = \sigma^r + \frac{E_s h_s^2 \Delta\kappa}{6 h_f (1-\upsilon_s)} \qquad 1$$

where $E_s$ and $v_s$ are the Young's modulus and the Poisson's ratio of the substrate respectively, $\sigma^r$ is the residual stress in the Si film due to sputter deposition, and $h_s$ and $h_f$ are the substrate and the film thickness respectively. Values for $E_s$ and $v_s$ were obtained from Brantley's work [8]. Residual stresses were measured by tracking the substrate curvature before and after each film-deposition (Ti/Cu, Si) step.

Based on *in situ* observations of height and volume changes during lithiation/delithiation in thin-film Si electrodes reported in the literature [9], it is reasonable to assume that the film thickness (also volume) increases linearly with SOC,

$$h_f = h_f^0 (1 + 2.7z) \qquad 2$$

where $h_f^0$ is initial film thickness, and $z$ represents the SOC, which can vary between 0 and 1. Here, z = 1 corresponds to a charge capacity of 3579 mAh/g, which corresponds to a volume expansion of 370%.

Substrate curvature was monitored with a multi-beam optical sensor (MOS) wafer curvature system (k-Space Associates, Dexter, MI), which is illustrated schematically in Figure 1. The MOS system uses a parallel array of laser beams that get reflected off the sample surface and captured on a camera. The relative change in the spot spacing is related to the wafer curvature through

$$\kappa = \frac{(d - d^0)}{d^0} \frac{1}{A_m} \qquad 3$$

where $d$ is the distance between two adjacent laser spots on the camera (see figure 1(b) in reference 3), $d_0$ is the initial distance and $A_m$ is the mirror constant, given by *2L/cos(θ)*; $L$ is the optical path length of the laser beam between the sample and the array and $\theta$ is the incident angle of the laser beam on the sample. The mirror constant $A_m$ is measured by placing a reference mirror of known curvature in the sample plane and measuring the relative change in the spot spacing. Data acquisition rate was 1 Hz for all the experiments.

At any SOC, when the film is delithiated by a small amount, the in-plane strain change ($\Delta\varepsilon$) is given by $\Delta\varepsilon = \Delta\sigma/Y + \Delta\varepsilon^*$, where $\sigma$ is the equi-biaxial stress in the film, $Y$ is the biaxial modulus of the film, $Y = E/(1-v)$, where $E$ is the Young's modulus and $v$ is the Poisson's ratio of the film; and $\varepsilon^*$ is the compositional strain due to lithiation. If we assume that the volumetric strain of the film due to lithiation is proportional to $z$, $\varepsilon^*$ is given by

$$\varepsilon^* = (1 + 2.7z)^{1/3} - 1 \qquad 4$$

Since the in-plane strain change is zero due to substrate constraint, *i.e.*, $\Delta\varepsilon = 0$, the biaxial modulus can be written as $Y = -\Delta\sigma/\Delta\varepsilon^*$. During the delithiation-relithiation perturbation at each SOC, the stress change $\Delta\sigma$ is obtained from the substrate-curvature change, and $\Delta\varepsilon^*$ is obtained from Eq. 4, from which the evolution of the biaxial modulus with SOC can be inferred.

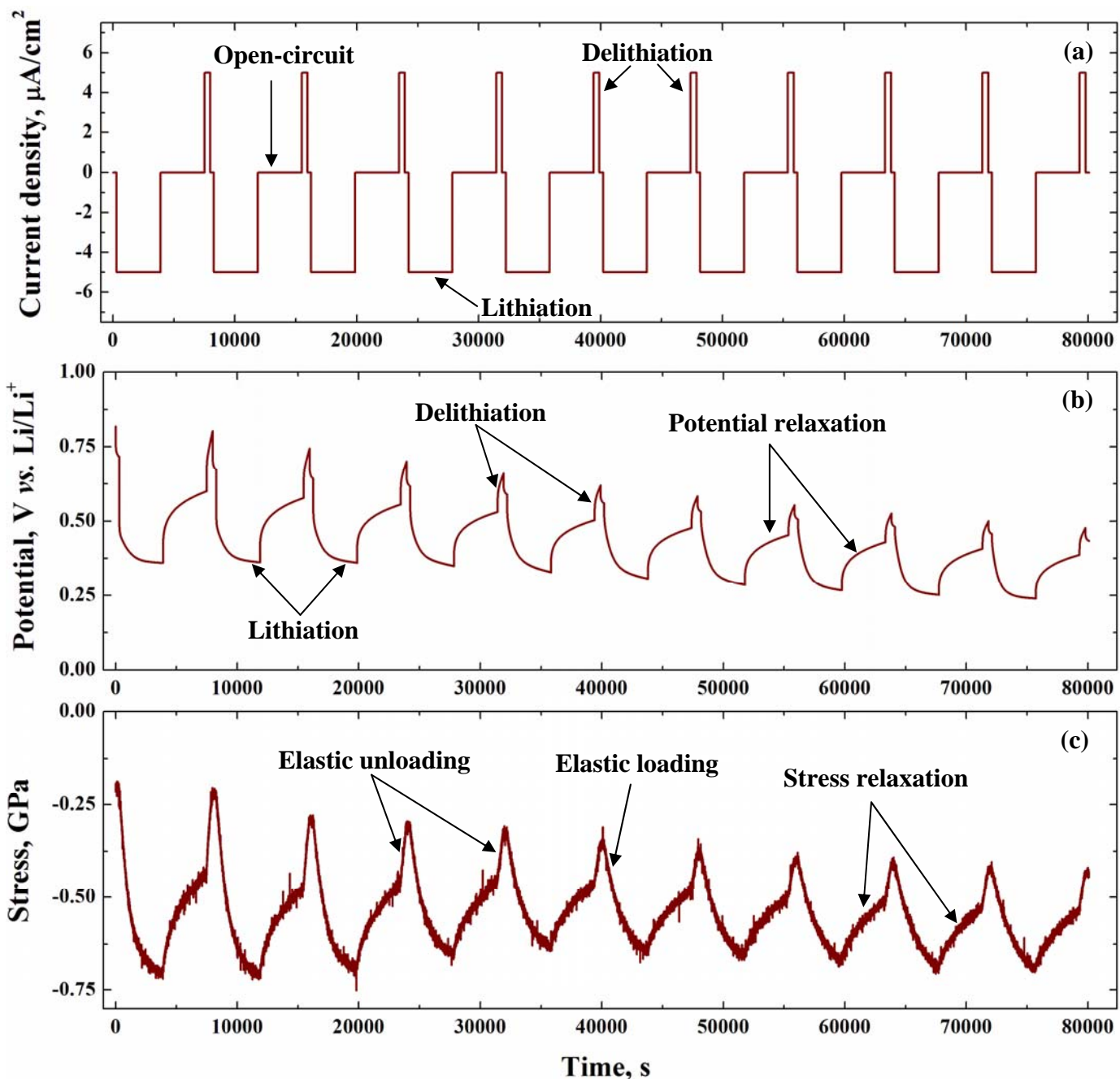

*Figure 2: Representative current-density, potential and stress transience obtained in situ from the experiments designed to measure biaxial moduli of a lithiated-silicon electrode.*

## 3. RESULTS AND DISCUSSION

It was shown earlier that the Li-Si electrodes undergo cycles of compressive and tensile stress during lithiation and delithiation, respectively [3]. Since full delithiation can result in large tensile stresses and film cracking, the elastic constants are best measured during the first lithiation, when the stress is known to be compressive and cracks do not form.

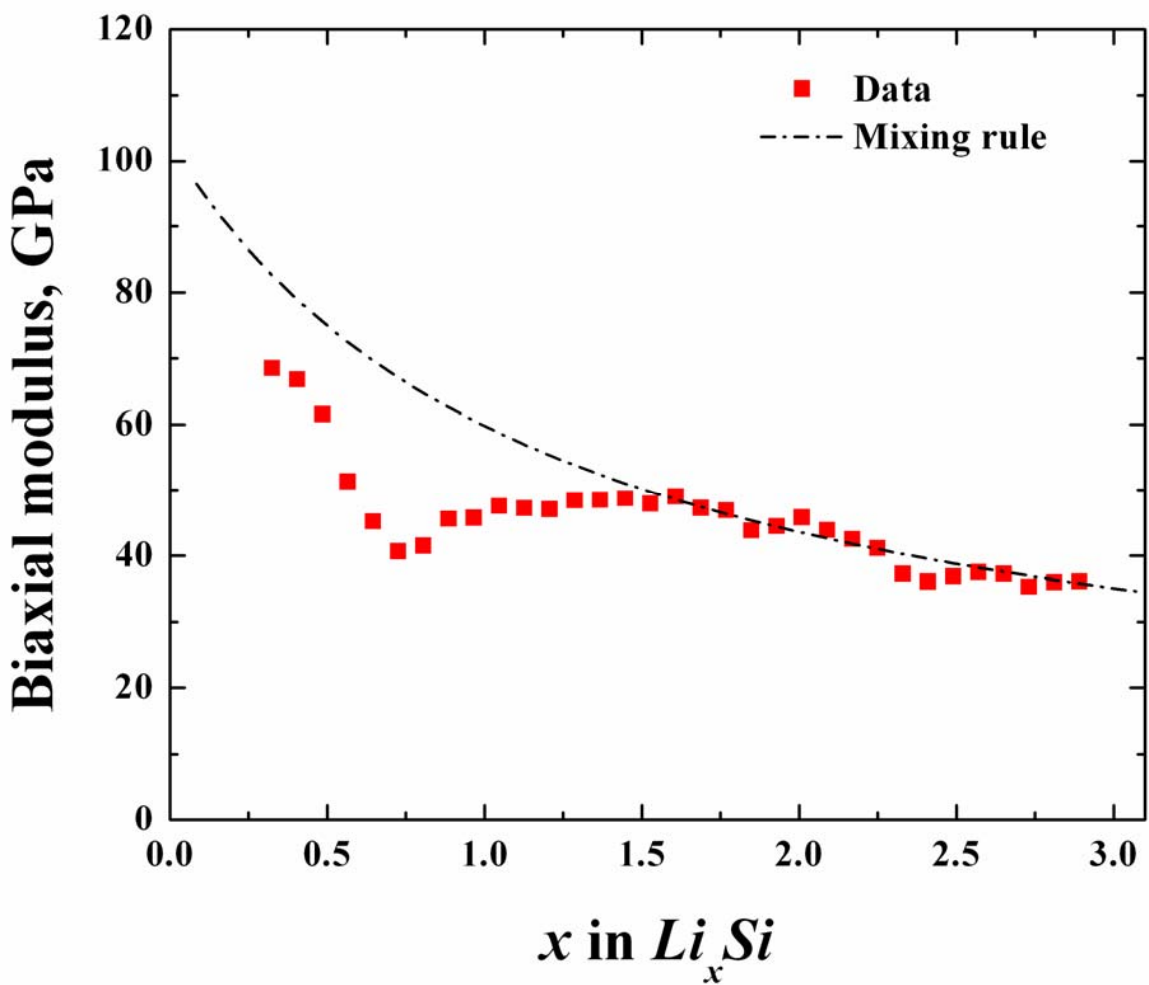

*Figure 3: Biaxial moduli of silicon thin-film as a function of lithium concentration obtained using data shown in Figure 2. The dotted line represents biaxial-modulus calculated using a linear mixing rule.*

A representative history of current-density, potential and stress (which is a fraction of a long sequence SOC perturbation cycles) is shown in *Figure 2*a-c. The stress-strain data in the elastic unloading region was used to estimate the biaxial modulus of the Li-Si system. Note the rapid stress relaxation immediately after the current interrupted, possibly due to double layer discharge. Ideally, the stress should reach a steady state during the open-circuit step before delithiation commences so that viscoplastic stress relaxation does not contribute to the stress drop. However, in the Li-Si system, side reactions slowly and continuously remove Li from Si, resulting in a corresponding continuous change in stress; hence the stress does not reach a steady-state value even after days [10]. In this investigation, we assume that one hour of open-circuit relaxation is sufficient to minimize the contribution of viscoplastic relaxation to the stress drop during delithiation.

The biaxial moduli of Li-Si calculated from the data set shown in *Figure 2*b-c, and the results are shown in *Figure 3*. The experiments reveal that the biaxial modulus decreases significantly, from around 70 GPa to about 35 GPa, as the Li fraction is increased. Hence, it becomes essential to consider the change in elastic moduli while modeling stress evolution and mechanical damage in Si anodes. We note that the DFT calculations of Shenoy *et al.* [4] agree well with our experimental measurements. For example, they show that the elastic constants of amorphous Li-Si are well described by a simple rule of mixtures between the properties of pure Si and pure Li. Using such a rule of a mixture, the biaxial modulus as a function of *x* can be estimated to be

$$Y_{est} = \frac{x_{Li}E_{Li} + x_{Si}E_{Si}}{1-(x_{Li}\nu_{Li} + x_{Si}\nu_{Si})} \qquad 5$$

where $x_{Li}(=1-x_{Si})$ is the fraction of Li atoms in the Li-Si electrode, $E_{Li}$, $E_{Si}$, and $\nu_{Li}$, $\nu_{Si}$, represent the Young's modulus and Poisson's ratio of amorphous-Si film and bulk Li, respectively. Parameters used for the calculations are listed in Table 1. The biaxial modulus estimates from Equation 5 are shown in *Figure 3* as a dotted line, which seems to agree with the experimental measurements remarkably well.

## 4. CONCLUSIONS

We report an *in situ* method to measure the change in the biaxial modulus of Si anodes as a function of Li concentration. It involves perturbing the SOC of the anode through a small delithiation-relithiation cycle, while measuring the change in film stress with a MOS wafer-curvature system. The measurements show that the biaxial modulus drops substantially from *ca.* 70 GPa for $Li_{0.32}Si$ to *ca.* 35 GPa for $Li_{3.0}Si$. A simple rule of mixtures was seen to agree well with the measurements. Hence, it may not be unreasonable to expect that other relevant mechanical properties (*e.g.*, ductility, fracture toughness) would also change substantially, which has implications for realistic modeling of the mechanics of Si anodes to predict their cycle life. The methodology presented here can be applied to characterize the biaxial modulus of other lithium-ion-battery electrodes of interest as well.

## 5. ACKNOWLEDGEMENTS

Support of this work by Brown's MRSEC sponsored by NSF under contract #DMR0520651; by the RI Science and Technology Advisory Council under grant #RIRA 2010-26; and by NASA grant # NNX10AN03A are gratefully acknowledged.